# Neutrino oscillometry at the next generation neutrino observatory

Yu.N. Novikov[1,2,*], T. Enqvist[3], A.N. Erykalov[1], F. v.Feilitzsch[4], J. Hissa[3], K. Loo[5],
D.A. Nesterenko[1,2], L. Oberauer[4], F. Thorne[4], W. Trzaska[5], J.D. Vergados[6], M. Wurm[7]

[1] *Petersburg Nuclear Physics Institute, Gatchina 188300, St.Petersburg, Russia*
[2] *St.Petersburg State University, Peterhof 198504, St.Petersburg, Russia*
[3] *University of Oulu, 90014 Oulu, Finland*
[4] *Technical University of München, 85748 München, Germany*
[5] *University of Jyväskylä, 40014 Jyväskylä, Finland*
[6] *University of Ioannina, 45110 Ioannina, Greece*
[7] *Universtity of Hamburg, 22761 Hamburg, Germany*

(Dated: September 29, 2011)

Abstract.  The large next generation liquid-scintillator detector LENA (Low Energy Neutrino Astronomy) offers an excellent opportunity for neutrino oscillometry. The characteristic spatial pattern of very low monoenergetic neutrino disappearance from artificial radioactive sources can be detected within the long length of detector. Sufficiently strong sources of more than 1 MCi activity can be produced at nuclear reactors. Oscillometry will provide a unique tool for precise determination of the mixing parameters for both active and sterile neutrinos within the broad mass region $\Delta m_{st}^2 \approx 0.01 - 2$ (eV)$^2$. LENA can be considered as a versatile tool for a careful investigation of neutrino oscillations.

## 1. Introduction

The discovery of neutrino oscillations, being one of the greatest triumphs of physics, can not answer many questions on the neutrino properties: How large is the neutrino absolute mass value? What can we say on the neutrino mass hierarchy? Are neutrinos Dirac or Majorana particles? Is the total lepton number conserved? Could we expect CP- as well as CPT-violation in the lepton sector? How many, if any, sterile neutrinos are there? etc. The answers to these questions will, perhaps, be given in future challenging experimental and theoretical investigations.

Meanwhile, the existence of oscillations demonstrates that the neutrino is a massive particle and should therefore be considered as beyond the simple Standard Model. We have already gotten a pretty good idea of the neutrino mixing matrix and we have two independent mass relations [1]:

$$\Delta m_{12}^2 = |m_1^2 - m_2^2| = (7.59 \pm 0.20) \times 10^{-5} (eV)^2 \tag{1}$$

and

$$\Delta m_{13}^2 \approx \Delta m_{23}^2 = |m_2^2 - m_3^2| = (2.45 \pm 0.09) \times 10^{-3} (eV)^2, \tag{2}$$

This means that the relevant L/E parameters are very different, which, in turn, disentangles the $L_{13}$ and $L_{12}$ neutrino oscillation lengths. As can be seen from (1) and (2) the global analysis led to a rather precise mass squared differences. Furthermore the two mixing angles $\theta_{12}$ and $\theta_{23}$ are well determined, whereas the mixing angle $\theta_{13}$ is still unknown. The last global analysis [2] gives nonzero $\theta_{13}$. It includes recent T2K [3], MINOS [4] and SNO-results [5].

* Corresponding author: novikov@pnpi.spb.ru

A recent careful analysis of Neutrino Anomaly (NA) [6,7] led to a challenging claim that this anomaly can be explained in terms of a new fourth neutrino, sterile in origin, with a much larger mass squared difference. Assuming that the neutrino mass eigenstates are non-degenerate, one finds [7]:

$$\Delta m^2_{14} \approx \Delta m^2_{24} = |m^2_2 - m^2_4| \approx 1.5 \text{ (eV)}^2 \quad (3)$$

with a mixing angle:
$$\sin^2(2\theta_{14}) = 0.14 \pm 0.08 \quad (95\% \text{ of c.l.}). \quad (4)$$

Similar results, $\Delta m^2_{14} \approx 2$ (eV)$^2$ and $0.01 \leq \sin^2(2\theta_{14}) \leq 0.13$, have been obtained in [6]. Also a value of $\Delta m^2_{14} \approx 5.6$ (eV)$^2$ has been discussed [8].

Subsequent analysis [9] showed, however, that existing data can be better explained, if two sterile neutrinos are included in the analysis. Their mass squared differences are estimated to be:
$$\Delta m^2_{14} = 0.47 \text{ (eV)}^2 \text{ and } \Delta m^2_{15} = 0.87 \text{ (eV)}^2. \quad (5)$$

Meanwhile, the analysis of [10] results in
$$\Delta m^2_{14} = 0.9 \text{ (eV)}^2 \text{ and } \Delta m^2_{15} = 1.61 \text{ (eV)}^2. \quad (6)$$

Even though these new neutrinos are sterile, i.e. they do not participate in week interaction, they should contribute to the oscillation, since they will tend to decrease the electron neutrino flux.

Fortunately the values for sterile neutrinos of (3) and (5) or (6) are strongly different from the values for active neutrinos (1) and (2). This disentangles oscillation lengths $L_{14}$ and $L_{15}$ from $L_{13}$ and $L_{12}$ and should allow the separation of the oscillation curves with and without sterile neutrinos.

In all previous experiments, the oscillation lengths $L_{13}$ and even more $L_{12}$ have been much larger than the size of detectors. For neutrinos with energies much higher than 1 MeV, the length $L_{13}$ is much longer than one kilometer. Obviously, it was unrealistic to have a detector of such long dimensions. Hence, all beam experiments, aiming at neutrino oscillations have so far considered just a single or at most a two-point measurement instead of the full oscillation curve.

A detector of a length of 100 m will become realistic in the near future. LENA is one of the advanced projects under discussion [11]. It opens a new unique possibility to scan the neutrino flavor changes, i.e. oscillations, within the dimensions of detector. This unprecedented possibility has been named "neutrino oscillometry" [12]. For successful oscillometry, the neutrino should be monoenergetic, and at rather small energy. Thus, an artificial radioactive source should be used. Such sources ($^{51}$Cr and $^{37}$Ar) have been used for calibration of the GALLEX [13] and SAGE [14] experiments aiming at solar neutrino investigations. $^{51}$Cr has been considered in the BOREXINO project [15]. The use of monoenergetic neutrinos has been proposed in the projects of neutrino oscillometry based on a spherical gaseous Time Projection Counter (TPC) in [12] and [16]. The use of $^{51}$Cr for investigations of the short-length neutrino oscillation is proposed also in [10] and [11].

In some specific cases, also antineutrino oscillometry can be performed both in a spherical TPC and with LENA [17,18]. The antineutrino spectral analysis can be used to search for sterile neutrinos in reactor experiments with two detectors placed very close to the nuclear reactor [19]. Recently, a short baseline oscillation experiment with a pion decay-at-rest beam has been proposed in [20].

In this article, we propose to use the giant Liquid Scintillator {LS} detector LENA for neutrino oscillometry, using a very strong nuclear electron capture source. The source will produce monoenergetic neutrinos at low energies, well suited to search for short-baseline neutrino oscillations via $\theta_{13}$ and into (multiple) sterile neutrinos, including a precise determination of the corresponding oscillation lengths. A short version of this work was given in the LENA white paper [11].

## 2. Neutrino oscillometry in a giant liquid scintillator detector

As it was mentioned above, the neutrino energy should correspond to an oscillation length contained within the detector, i.e. much less than 1 MeV in the case of $L_{13}$. At these energies, neutrinos can only be detected via the elastic scattering on electrons of the target [21]. However the cross-section of this reaction is very small. In fact, Fig. 1 (see, e.g., [12]) shows that the cross-section for 100 keV neutrinos is equal to $2\times10^{-46}$ cm$^2$. Therefore, a compensation of this smallness is needed. Keeping in mind these basic requirements, we can formulate the general conditions for the oscillometry method, which is point-by-point measurement of the recoil electron numbers along the neutrino path within the LS-detector. They are as follows:

- In order to have a single oscillation curve, the neutrinos should be monoenergetic, while their energy has to be less than 1 MeV in order to get oscillations within the detector. At the same time, the energy should be higher than 200 keV, which is the energy threshold for the liquid scintillator [22].
- Since the neutrino-electron cross section is tiny, this smallness should be compensated by other detector parameters: a large target volume, high source intensity on the level of a few MCi, etc.
- The lifetime of the source should be not too long in order to provide the maximum possible intensity of neutrinos escaped from the nucleus after electron capture, which dominates the radioactive decay of nuclides in the source.
- It is possible to produce such monoenergetic neutrino emitters in sufficient quantities in long-term irradiations of the corresponding stable nuclides at nuclear reactors. These sources are suitable for multiple use and irradiations, since the source can be installed outside, apparently at the top of cylindrical detector.

The determination of the mixing angle $\theta_{13}$ or the exploration of new (sterile) neutrinos can be done experimentally by the direct measurements of the oscillation curves from monoenergetic neutrino-electron scattering.

### 2.1 Oscillation probability

We can apply a four or more neutrino oscillation analysis to write, under the approximations of Eqs. (1) – (5), the $\nu_e$ survival probability as follows:

$$P(\nu_e \to \nu_e) \approx 1 - \chi(E_\nu)\left[\sin^2 2\theta_{12}\sin^2\left(\pi\frac{L}{L_{12}}\right) + \sin^2 2\theta_{13}\sin^2\left(\pi\frac{L}{L_{13}}\right)\right] \quad (7)$$

$$- \sum_{n=3}^{24}\sin^2 2\theta_{1n}\sin^2\left(\pi\frac{L}{L_{1n}}\right)$$

with

$$L_{ij} = \frac{4\pi E_\nu}{m_i^2 - m_j^2} . \quad (8)$$

The length $L_{ij}$ in meters can be written to

$$L_{ij}[m] = \frac{2.48 \cdot E_\nu[\text{MeV}]}{\Delta m_{ij}^2[\text{eV}^2]} . \quad (9)$$

Since the oscillation lengths are very different, $L_{14}$ and $L_{15} \ll L_{13} \ll L_{12}$, one may select the distance L so that one observes only one mode of oscillation, e.g. into sterile neutrinos.

Let's consider separately the short-length oscillation properties with and without sterile neutrinos:

**a)** In the three-flavor scenario, the electron neutrino, produced in weak interactions, can be expressed in terms of a linear superposition of the three mass eigenstates:

$$\nu_e = \cos\theta_{12}\cos\theta_{13}\nu_1 + \cos\theta_{12}\cos\theta_{13}\nu_2 + \sin\theta_{13}e^{i\delta}\nu_3. \tag{10}$$

Therefore, the $\nu_e$ survival probability is given by [12]:

$$P(\nu_e \to \nu_e) = \tag{11}$$
$$1 - \left[\cos^4\theta_{13}\sin^2 2\theta_{12}\sin^2\left(\pi\frac{L}{L_{12}}\right) + \sin^2\theta_{12}\sin^2 2\theta_{13}\sin^2\left(\pi\frac{L}{L_{23}}\right)\right.$$
$$\left. + \cos^2\theta_{12}\sin^2 2\theta_{13}\sin^2\left(\pi\frac{L}{L_{13}}\right)\right],$$

$\sin 2\theta_{13}$ is a small quantity [2] and it will be further investigated below. In the limit of $\cos^4\theta_{13} \to 1$, one recovers the standard expression familiar from the solar neutrino oscillation analysis, if we take $\sin\theta_{12}=\sin\theta_{solar}$, $\cos\theta_{12}=\cos\theta_{solar}$, $\theta_{solar}$ as determined from the solar neutrino data.

Assuming $\Delta m_{13}^2 \approx \Delta m_{23}^2$ and $L_{13} \approx L_{23}$ we find:

$$P(\nu_e \to \nu_e) \approx 1 - \left[\sin^2(2\theta_{solar})\sin^2\left(\pi\frac{L}{L_{12}}\right) + \sin^2(2\theta_{13})\sin^2\left(\pi\frac{L}{L_{13}}\right)\right], \tag{12}$$

in which the term containing $\sin^2(2\theta_{13})$ and the short oscillation length $L_{13}$ ($L_{13} \approx \frac{1}{33} L_{12}$) is relevant for the experimental approach discussed in the present paper. In addition to the signal generated by the surviving electron neutrinos, there are also electron recoils induced by neutral current interactions of the two other active neutrino flavors [21]. These flavors are generated via the appearance oscillation:

$$P\left(\nu_e \to \sum_{\alpha \ne e} \nu_\alpha\right) \approx \left[\sin^2(2\theta_{solar})\sin^2\left(\pi\frac{L}{L_{12}}\right) + \sin^2(2\theta_{13})\sin^2\left(\pi\frac{L}{L_{13}}\right)\right]. \tag{13}$$

Thus, the number of the recoil electrons of energy $T$ is proportional to the ($\nu_e$, e−) scattering cross section with a proportionality constant $C(E_\nu, T)$ given by:

$$C(E_\nu,T) = 1 - \chi(E_\nu,T)\left\{\sin^2(2\theta_{solar})\sin^2\left(\pi\frac{L}{L_{12}}\right) + \sin^2(2\theta_{13})\sin^2\left(\pi\frac{L}{L_{13}}\right)\right\}. \tag{14}$$

The function $\chi(E_\nu,T)$, which represents the relative difference between the cross sections $\nu_e$ and $\nu_{\mu,\tau}$, is discussed in [21]. Note, however, that the cross sections of $\nu_{\mu,\tau}$ are much smaller than that of $\nu_e$, as shown in the Fig. 1.

Assuming the value $\Delta m_{13}^2 = 2.45 \times 10^{-3} eV^2$ for the mass squared difference that can be derived from the global oscillation analysis [1], according to (9) the oscillation length expressed in meters is approximately equal to the neutrino energy in keV:

$$L_{13}[m] \approx E_\nu[\text{keV}], \tag{15}$$

which is about 3% of the solar oscillation length $L_{12}$.
Thus, the oscillation length e. g. for 100 keV neutrinos is 100 m.

**b)** For a 3+1 or 3+2 scenario including sterile neutrinos, the shortest baseline is either due to $\Delta m_{14}^2 = 1.5\ eV^2$ [7] or to $\Delta m_{14}^2 = 0.47\ eV^2$ and $\Delta m_{15}^2 = 0.87\ eV^2$ [9]. In the 3+1 case, the oscillation length $L_{14}$ for a EC source experiment is $\approx 1$ m. The $L_{14}$ and $L_{15}$ values in the 3+2 scenario [9] are, respectively, 3.94 m and 2.13 m. Since the L-values for both scenarios are on a comparable level of approximately 1 m, we restrict the analysis below to 3+1.

The survival probability of electron neutrinos (see eq. 7) in a short baseline experiment can be approximated as

$$P(\nu_e \to \nu_e) \approx 1 - \left[\sin^2 2\theta_{14} \sin^2\left(\pi \frac{L}{L_{14}}\right)\right] \quad (16)$$

as long as the mixing effects are clearly disentangled from ative neutrino oscillations due to the shorter oscillation length $L_{1j}$.

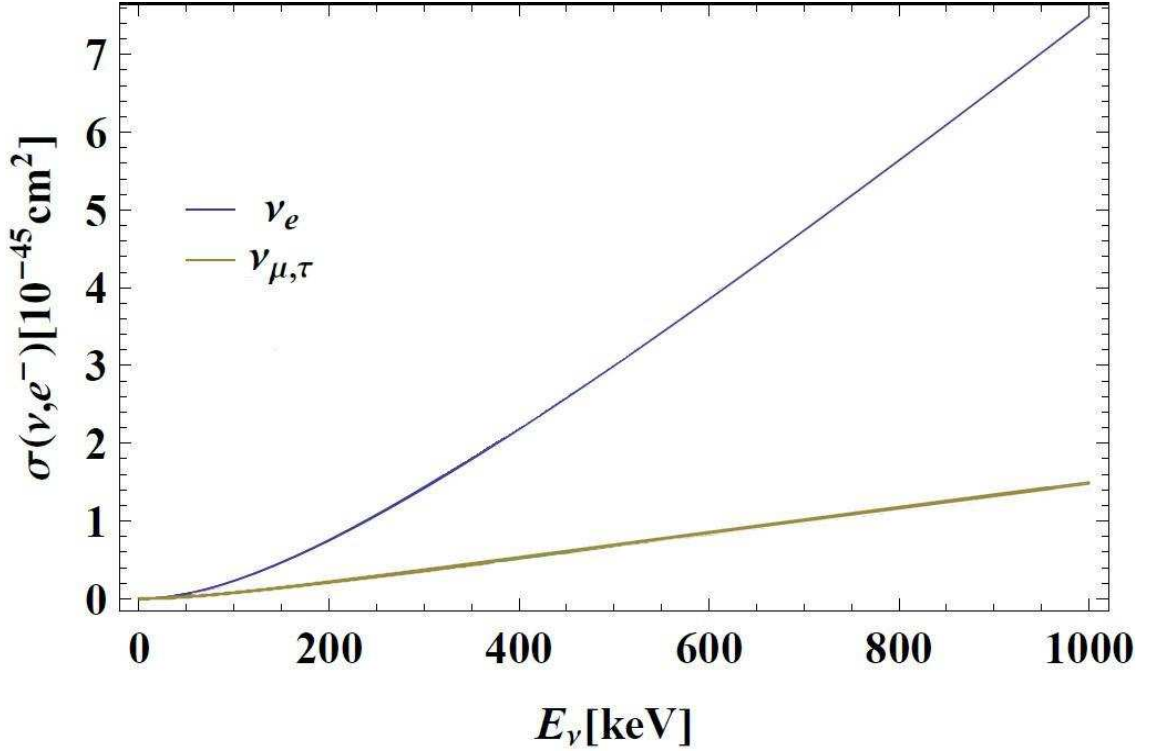

Fig. 1. The total ($\nu_e$, e−) and ($\nu_\mu$, e−) cross sections in the absence of oscillations as a function of the neutrino energy [12]. In the present work, the neutrinos have a definite energy.

## 2.2 Neutrino-electron cross section

For low energy neutrinos, the differential cross section [23] takes the form

$$\frac{d\sigma}{dT} = \left(\frac{d\sigma}{dT}\right)_{weak} + \left(\frac{d\sigma}{dT}\right)_{EM} . \quad (17)$$

The second term, due to the neutrino magnetic moment, is inversely proportional to the recoil electron energy and, at low electron energies, may be used to considerably improve the current limit of the neutrino magnetic moment. However, this is not the subject of the present study. The cross section in the rest frame of the initial electron due to weak interaction alone becomes [21,12]:

$$\left(\frac{d\sigma}{dT}\right)_{weak} = \frac{G_F^2 m_e}{2\pi}\left[(g_V + g_A)^2 + (g_V - g_A)^2(1 - \frac{T}{E_\nu})^2 + (g_A^2 - g_V^2)\frac{m_e T}{E_\nu^2}\right] \quad (18)$$

$g_V = \sin^2(2\theta_W) + 1/2, g_A = 1/2$ for $\nu_e$,
$g_V = \sin^2(2\theta_W) - 1/2, g_A = -1/2$ for $\nu_\mu, \nu_\tau$,
with $\theta_W$ – the Weinberg angle.
For antineutrinos $g_A =\to -g_A$.
The scale is set by the week interaction:

$$\frac{G_F^2 m_e}{2\pi} = 4.45 \times 10^{-48} \frac{cm^2}{keV} . \quad (19)$$

The recoil electron energy depends on the neutrino energy and the scattering angle. It is given by:

$$T = \frac{2m_e(E_\nu \cos\theta)^2}{(m_e + E_\nu)^2 - (E_\nu \cos\theta)^2} \quad . \tag{20}$$

The total cross sections as a function of neutrino energy for electron, muon and tau neutrino are shown in Fig. 1.

## 3. Detection principle

Liquid-scintillator detectors are well-suited for the detection of sub-MeV neutrinos via elastic neutrino-electron scattering. As the cross section for this interaction is tiny, one needs to compensate this smallness by the largeness of the interaction volume, the density of the target material, the strength of the neutrino source, the measurement time and other less important parameters. Below, we discuss shortly the influence of all these components to the differential number of scattering events within the detector which will compose the oscillometry curve.

### 3.1 Design of detector

The artist's view of detector taken from [11] is shown in Fig. 2. The cylindrical detector tank encloses a large volume of 100 m height and 30 m diameter. It will be filled with an LAB-based liquid scintillator, featuring a density of 863 kg/m$^3$, corresponding to an electron number density of $3 \times 10^{29}$ e/m$^3$. The expected fiducial volume for sub-MeV neutrino detection is ≈ 35 kt.

Detection of the scintillation light will be done by about 50,000 photomultipliers surrounding the detector body. Assuming a light yield of 200 pe/MeV, the expected position sensitivity should be ≈ 0.25 m at 500 keV electron recoil energy. In this energy region, the energy resolution will be ≈ 10% [11].

Due to the large number of carbon atoms contained in the scintillator, even the relatively small amount of residual $^{14}$C will effectively mask recoil electrons at energies below 200 keV, because the endpoint of the continuous β-spectrum of $^{14}$C is 156 keV. Therefore, the calculations of oscillation spectra have to take into account this intrinsic electron threshold energy $T_{th}$.

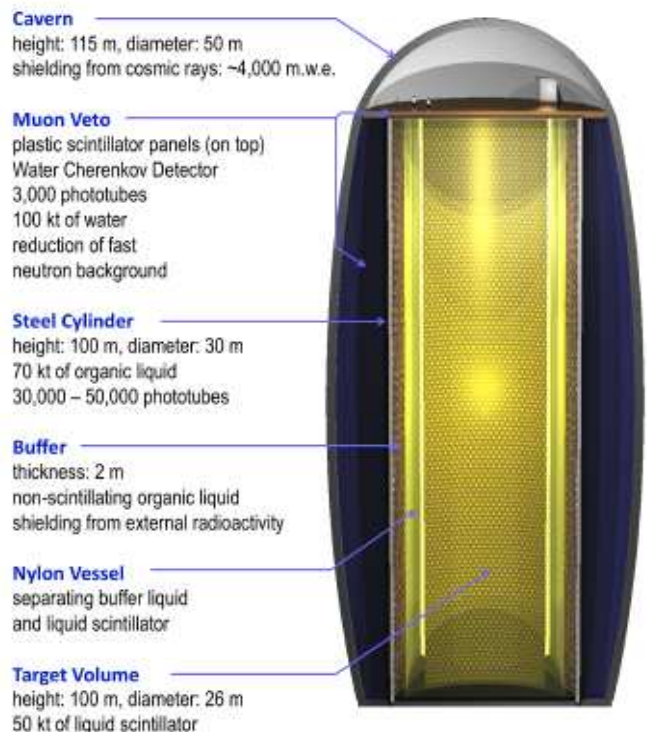

Fig. 2. Schematic view of the proposed liquid-scintillator detector LENA [11].

## 3.2 Energy threshold for recoil electrons

The threshold effect on (ν,e)-yield can be calculated analytically.
The total neutrino electron scattering yield for the active neutrinos can be cast in the form:

$$\sigma(L, x, y_{th}) = \sigma(0, x, y_{th}) \cdot (1 - \chi(x, y_{th}) p(L, x)) \tag{21}$$

with

$$p(L, x) = \sin^2\left(\pi \frac{\Delta m_{12}^2 \, [eV^2]}{2.48 m_e \, [MeV]} \cdot \frac{L \, [m]}{x}\right) \cdot \sin^2(2\theta_{solar}) \tag{22}$$
$$+ \sin^2\left(\pi \frac{\Delta m_{13}^2 \, [eV^2]}{2.48 m_e \, [MeV]} \cdot \frac{L \, [m]}{x}\right) \cdot \sin^2(2\theta_{13})$$

$L$ is the distance between source and detector, $x = E_\nu/m_e$, $m_e$ is the electron mass and $y_{th} = (T_e)_{th}/m_e$ with $(T_e)_{th}$ is the electron threshold energy imposed by the detector. We find

$$\sigma(0, x, y_{th}) = \frac{G_F^2 m_e^2}{2\pi} \cdot \frac{1}{x^2(2x+1)^3} \cdot [(x(17.7x + 15.3) + 3.35)x^4 - 18.8(x + 0.5)^3 y_{th} x^2 \tag{23}$$
$$- 0.569(x + 0.5)^3 y_{th}^3 + 1.71(x + 0.5)(x + 1.58)(x(x + 1) + 0.25) y_{th}^2]$$

and

$$\chi = 1 - N/D \tag{24}$$

$$N = \frac{4((4x + 3)^2 + 0.304x(4x + 3) + 0.006(4x(4x + 3) + 3))x^4}{(2x + 1)^3} \quad 12.1 y_{th} x^2 \quad 1.71 y_{th}^3 \tag{25}$$
$$+ 2.77(2x - 0.038(4x + 2) - 1) y_{th}^2$$

$$D = \frac{4((4x + 3)^2 + 7.70x(4x + 3) + 3.70(4x(4x + 3) + 3))x^4}{(2x + 1)^3} - 56.4 y_{th} x^2 - 1.71 y_{th}^3 \tag{26}$$
$$+ 2.77(-2x + 0.962(4x + 2) + 1) y_{th}^2$$

For zero threshold, these reduce to the formulas:

$$\sigma(0, x) = \frac{G_F^2 m_e^2}{2\pi} \cdot \frac{x^2(17.746 x^2 + 15.310 x + 3.362)}{(2x + 1)^3}, \tag{27}$$

which is the total cross section in the absence of oscillations. Furthermore

$$\chi(x) = \frac{0.551496(2x + 1)(4x + 1)}{5.27285 x^2 + 4.5515 x + 1}, \tag{28}$$

We assume that the volume of the source is much smaller than the volume of the target/detector.

In case of sterile neutrinos, their existence is indicated by the rate reduction due to the disappearance of $\nu_e$ via the mixing angle $\theta_{14}$. Thus, the number of the scattered electrons is proportional to the ($\nu_e$, e⁻) scattering yield, which can be given in the form:

$$\sigma(L, x, y_{th}) = \sigma(0, x, y_{th}) \cdot (1 - p(L, x)). \tag{29}$$

In this case, χ(x) is unity [21,12].
The oscillation probability induced by sterile neutrinos takes the form:

$$p(L, x) = \sin^2\left(\pi \frac{\Delta m_{14}^2 \, [eV^2]}{2.48 m_e \, [MeV]} \cdot \frac{L \, [m]}{x}\right) \cdot \sin^2(2\theta_{14}). \tag{30}$$

## 3.3 Geometrical factor for LENA detector

Another specific problem occurs if we install the neutrino source on the top of the cylinder. In this case, a "geometrical factor" has to be taken into account for calculating the event rate as a function of the distance $L$ to the neutrino source. Due to the isotropic emission of the neutrinos, there will be a decrease in rate due to the effective acceptance angle provided by the detector. This will be especially true for large $L$, in which case the cone with the vertex at the place of source will be smaller than a half sphere and the rate will decrease approximately as $1/L^2$.

In the following, $z$ will describe the distance from the neutrino source along the detector axis, while $r$ describes the radial distance to the axis. The differential number of recoil electron events between $r$ and $r + dr$ and $z$ and $z + dz$ is given by:

$$dN = N_\nu n_e \frac{2\pi r\, dr\, dz}{4\pi(r^2 + z^2)} \sigma\left(\sqrt{r^2 + z^2}, x, y_{th}\right). \tag{31}$$

The differential number can also be written as a function of the distance to the source $L$, transforming from the variables $(r, z)$ to $(L, \zeta)$ and integrating over the angle $\zeta$. It yields:

$$\frac{dN}{dL} = \frac{1}{2} N_\nu n_e g_{av}(L, R_0, H) \sigma(L, x, y_{th}), \tag{32}$$

where $g_{av}(L, R_0, H)$ is the geometrical factor depending on $L$; $R_0$ and $H$ are radius and height of the fiducial volume. The geometrical factor can be cast in the form:

$$g_{av}(L, R_0, H) = \begin{cases} 1, & 0 < L < R_0 \\ 1 - \dfrac{\sqrt{L^2 - R_0^2}}{L}, & R_0 < L < H \\ \dfrac{H}{L} - \dfrac{\sqrt{L^2 - R_0^2}}{L}, & H < L < \sqrt{R_0^2 + H^2}. \end{cases} \tag{33}$$

For $R_0 = 11$m and $H = 90$m, $g_{av}(L, R_0, H)$ is shown in Fig. 3. To disentangle the oscillation dependence $L$ from geometrical effects, the observed rate must be divided by $g_{av}$. The resulting curve for $^{51}$Cr is shown in Fig. 4.

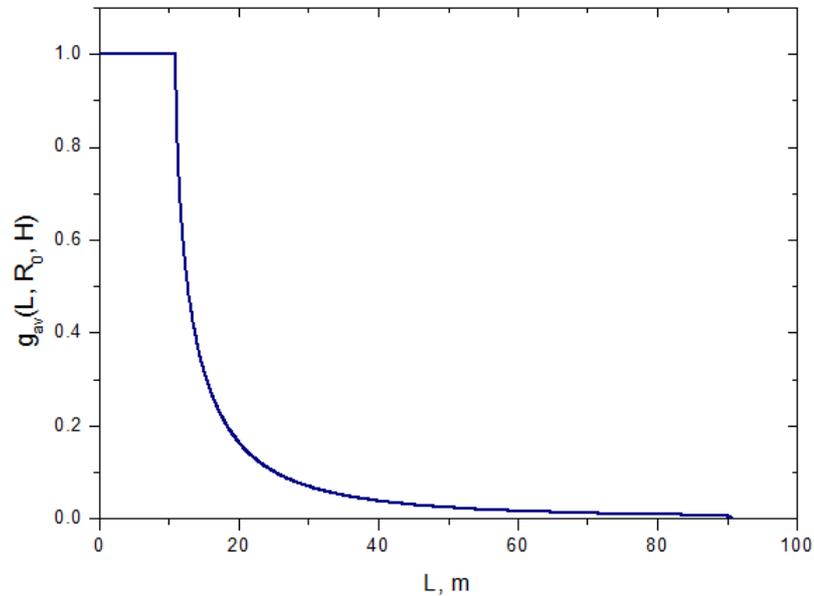

Fig. 3. The geometrical factor $g_{av}(L, R_0, H)$ as a function of $L$ for the value $R_0 = 11$m, $H = 90$m. Note the flux factor of 1/2 has explicitly put in Eq. (33) and thus it is not included in the geometry. In the first region, which is a spherical segment, one sees that the geometrical factor is unity like the case of the sphere.

The integral number of events $N_{int}$ can be obtained by integration of eq. (32):

$$N_{int} = N_0[1 - a(L_{1i}, H, R_0) \cdot \sin^2(2\theta_{1i})], \tag{34}$$

where $N_0$ is the expected event number without oscillations, while the fraction of "disappearing" neutrino events is a function of the oscillation probability. Eq. (34) can be used as basis for estimating the sensitivity for the detection of the neutrino $1i$-mixing ($i \geq 3$).

## 4. Monoenergetic neutrino sources

As mentioned above, the neutrinos should have a low energy and should be monoenergetic, i.e. with a defined $L_{1i}$-value. The source intensity should be very high. These requirements for the source characteristics are very severe. However, a screening of the available data bases results in a number of suitable radioisotopes undergoing electron capture, thereby emitting sub-MeV neutrinos.

Since this process is characterized by two-body final states, the total neutrino energy is equal to the difference of the total capture energy $Q_{EC}$ and the sum of binding energy of captured electron $B_j$ and the γ-de-excitation energy of daughter nuclide, $E_\gamma$. The nuclear recoil energy can be neglected.

$$Q_\nu = Q_{EC} - (B_j + E_\gamma) \quad \text{(with } j \equiv K, L, .. \text{ atomic electron orbits).} \tag{35}$$

The value of $Q_\nu$ can be easily determined because the capture energies are usually known (or can be measured very precisely by ion-trap spectrometry [24]) and the electron binding energies are tabulated with high precision [25]. The main feature of the electron capture process is the monochromaticity of the emitted neutrinos. As we mentioned above, just this paves the way for the neutrino oscillometry.

In order to provide the required high source intensity, the half-life of the isotope should not be too long. At the same time, it should not be too short as in this case both the $Q_{EC}$-values and consequently the neutrino energies $E_\nu \approx Q_\nu$ are rather high. In this case, $L_{13}$ for active neutrinos will exceed the detector length considerably.

When choosing the isotope, also the recoil electron detection threshold in liquid scintillator has to be taken into account. A value of of 200 keV has been established by the BOREXINO-experiment [26]. Therefore, the optimum neutrino energy appropriate for LENA should be several hundreds of keV.

Based on these considerations, we collected possible candidates for neutrino oscillometry. They are presented in Table 1, which contains only nuclides that can be produced by irradiation of lighter (A-1) stable isotopes of the same chemical element (shown in column 6) at a nuclear reactor. Column 4 shows only the main components of neutrino emission for each nuclide. The information for other weak components is given in Table 2 for $^{51}$Cr. This nuclide was used in the past for calibration of solar neutrino detectors [13]. A $^{51}$Cr source of an initial activity of 1.7 MCi [27] was produced by placing 36 kg of metallic chromium, enriched in ≈ 40% of $^{50}$Cr, at the core of the Siloe reactor in Grenoble (35MW thermal power) for a period of 23.8 days. In principle, the 36 kg batch of enriched chromium is still available and could be reused for LENA. Table 2 shows all neutrino branches which occur in the decay of $^{51}$Cr. The electron capture in $^{51}$Cr populates the ground state of $^{51}$V with a branching ratio of 90%, while it decays with a probability of 10% to an excited state at 320 keV. Two branches to both nuclear ground and excited states with the electron capture from K- and L-orbits are shown in Table 2 with their intensities. The weak part of branches with the neutrino energies of 426.5 keV and 421.4 keV and with the sum intensity of 10% will contribute to the oscillometry curve by the 2.5 times weaker intensity because of 2.5 times smaller neutrino-electron cross sections for this neutrino energy region in comparison to 746 keV neutrinos (see Fig. 1). Thus, the contribution of neutrino branches with the length $L_{13} \approx 426$ m to the oscillometry curve is about 4 %. It can be fitted and considered as a background. As it can be seen from the last column of Table 1 we can obtain about 10 MCi for $^{51}$Cr taking into account the available weight of $^{50}$Cr and irradiation time at the nuclear reactor.

**Table 1.** Potential EC $\nu_e$ sources that can be produced by neutron irradiation in nuclear reactors. The half-life $T_{1/2}$, the Q-value of the Electron Capture, the energy $E_\nu$ of the neutrino and the corresponding branching ratio BR, as well as the maximum electron recoil energy $E_{e,max}$ are shown. The achievable neutrino source intensities have been estimated for 1 kg batches of the irradiated elements, assuming natural isotope abundances (in exclusion of $^{50}$Cr) and a 10-day irradiation with a neutron flux of $5 \cdot 10^{14}$ n/cm$^2$s. Neutron-capture cross sections were taken from [28].

| Nuclide | $T_{1/2}$ [d] | $Q_{EC}$ [keV] | $E_\nu$ [keV] | $E_{e,\,max}$ [keV] | Target material | $\nu$-intensity [kg$^{-1}$s$^{-1}$] |
|---|---|---|---|---|---|---|
| $^{37}$Ar | 35 | 814 | 811 (100%) | 617 | $^{40}$Ca, Ar | 8.3x10$^{15}$ |
| $^{51}$Cr | 28 | 753 | 747 (90%) | 560 | $^{50}$Cr | 2.3x10$^{16}$ |
| $^{75}$Se | 120 | 863 | 450 (96%) | 287 | Se | 1.1x10$^{14}$ |
| $^{113}$Sn | 116 | 1037 | 617 (98%) | 436 | Sn | 8x10$^{11}$ |
| $^{145}$Sm | 340 | 616 | 510 (91%) | 340 | Sm | 2x10$^{12}$ |
| $^{169}$Yb | 32 | 910 | 470 (83%) | 304 | Yb | 1.1x10$^{15}$ |

**Table 2.** Monoenergetic neutrino branches in the electron capture by $^{51}$Cr. The K/L capture branching ratio for $^{51}$Cr is 11 [29]

| $Q_{EC}$ [keV] | $B_j$ (K,L) [keV] | $E_\nu$ [keV] | Branching ratio [%] | Contribution to $L_{1i}$–curve [%] |
|---|---|---|---|---|
| 752 | 5.5 (K) | 746.5 | 82 | 88 |
|  | 0.6 (L) | 751.4 | 8 | 8 |
| 422 | 5.5 (K) | 426.5 | 9 | 3.6 |
|  | 0.6 (L) | 421.4 | 1 | 0.4 |

Table 2 shows that the main contribution of 96% to $L_{1i}$ originates from the neutrino energy doublet of 746.5 keV and 751.4 keV, in which the major part belongs to the first line. All other contributions can be considered as a background: they are well known and can be fitted.

## 5. Physics case for monoenergetic neutrino oscillometry

The main goal of oscillometry is the determination of the oscillometry curve, i.e. the oscillation length $L_{1i}$ and mixing angles $\theta_{1i}$. Global best fit parameters [1,2] suggest that the different $L_{1i}$-values are disentangled. For a neutrino energy of 747 keV ($^{51}$Cr), the oscillation baselines are 22 km for $L_{12}$, 740 m for $L_{13}$, and $\leq 4$ m for $L_{14}$ and $L_{15}$. Thus, LENA can be used to investigate neutrino oscillations on the scale of $L_{13}$ and $L_{14}/L_{15}$.

### 5.1 Mixing parameters $\theta_{13}$ and $L_{13}$ for active neutrinos

An advantage of the short-baseline oscillometry for a precise determination of $\theta_{13}$ is the absence of matter effects. In long-baseline neutrino beam experiments, these effects cause a degeneracy [30] in the determination of the oscillation parameters. The oscillometric approach to determining $\theta_{13}$ is a measurement of the differential rate d$N$/d$L$ of $\nu$-e scattering events as a function of the distance $L$ from the neutrino source (see equation (32)). Due to the still relatively large oscillation length $L_{13}$, e.g., 740 m for $^{51}$Cr, such a measurement requires not only a strong

EC source but also repeated measurements with several strong sources. Fig. 4 shows d$N$/d$L$ for the case of a 5x55 days measurement campaign at LENA, based on a $^{51}$Cr source of an initial activity of 5 MCi, placed at the top of the detector. A detection threshold of 200 keV is assumed. The rates have been normalized to the full solid angle. The dashed lines indicate the statistical 1σ uncertainties on the differential rate, assuming a bin width of 10 m: For large L, these uncertainties increase substantially due to the geometric decrease in the detected event rate with $L^2$.

Alternatively, the mixing angle can be determined by observing a deficit in the integral number of events via Eq. (34). For a spherical detector, the event rate in units of Λ will take the form:

$$N/\Lambda = -A\sin^2 2\theta_{13} + B. \tag{36}$$

$$\Lambda = \frac{G_F^2 m_e^2}{2\pi} R_0 N_\nu n_e, \tag{37}$$

with $N_\nu$ the number of neutrinos emitted by the source, $n_e$ the density of electrons in the target, $R_0$ the radius of the target, $G_F$ the Fermi constant and $m_e$ the electron mass. For a cylindrical detector, this parameterization has to be multiplied by a geometric factor which depends on the ratio $R_0/H$. For LENA ($R_0 = 11$m, $H = 91$m), it is approximately 0.75.

Thus, the total number of events can be expressed in the simple way which can be used for the test of sensitivity determination of $\theta_{13}$:

$$N_0 = -C\sin^2 2\theta_{13} + D. \tag{38}$$

Rough estimations taking into account the geometric factor of equation (33), an energy threshold $(T_e)_{th} = 200$ keV, 5x55 days of measurement and an initial activity of 5 MCi result in $C=2.9\times10^4$ and $D=7.0\times10^6$ for $^{51}$Cr. $C=4.8\times10^4$ and $D=4.6\times10^6$ are obtained in the case of a $^{75}$Se source featuring an activity of 3.5 MCi, a lower neutrino energy $E_\nu = 450$ keV and therefore a shorter oscillation baseline $L_{13} \approx 450$ m (see Table 1).

Fig. 5 shows 90% exclusion limits for the observation of $\nu_e$ disappearance for both nuclides. $^{75}$Se reaches by far better limits due to the shorter baseline and therefore larger value of $a(L_{1i}, H, R_0)$ (in eq. 34). A sensitivity of $\sin^2(2\theta_{13}) \approx 0.1$ could be achieved by five runs (of 168 days each) with a 3.5 MCi selenium source, reaching the currently indicated range of interest [2]. However, the required exposure will further increase when uncertainties introduced by the subtraction or suppression of backgrounds are considered [11]. Neglecting backgrounds, shifting the source to larger distances from the detector will increase the sensitivity as the oscillation becomes more prominent for larger values of $L$ [31]. Sensitivity might also be improved by lowering the detection threshold below ≈200 keV (cf. BOREXINO [26]), which could be achieved by a significant reduction of the 14C concentration in the scintillator material.

In principle, also the oscillation length $L_{13}$ could be determined by oscillometry, although even for $^{75}$Se, the spatial oscillation pattern is only partially contained within the extensions of LENA. The result could be compared with the neutrino energy that is usually well known or that can be measured independently very precisely [24]. For $^{51}$Cr, the neutrino energy is presently known with a precision of 0.03%. Since eq. (9) is valid if the global-analysis value of $\Delta m_{13}^2 = 2.45 \times 10^{-3}$ eV$^2$ is used, this comparison will be helpful for assessment of the global analysis itself.

In summary, the sensitivity for $\sin^2(2\theta_{13})$ that could be achieved in LENA based on strong EC sources and multiple exposures is of the order of current reactor experiments like Double Chooz, T2K, RENO and Daya Bay (see, e.g., [32]), but does not exceed it. However, it will provide a complementary method to determine $\theta_{13}$ and $L_{13}$ provided $\sin^2(2\theta_{13})$ is close to the current best fit value [2].

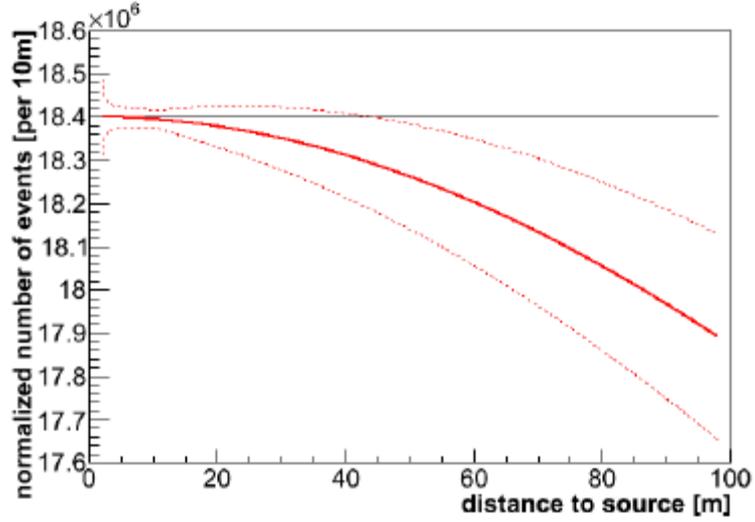

Fig. 4. Differential ν-e scattering event rate for a measurement campaign based on a $^{51}$Cr source installed on top of LENA (5×55 d, 5 MCi). The dashed lines indicate the statistical uncertainties (1σ) assuming a bin width of 10 m. The assumed oscillation amplitude concerns $sin^2(2\theta_{13}) = 0.17$.

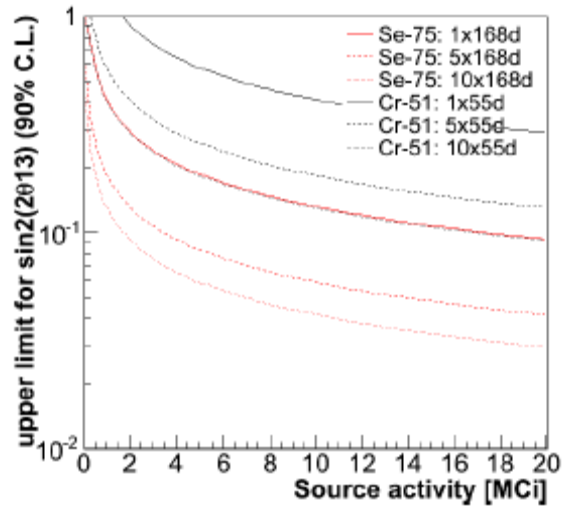

Fig. 5. Upper limits for $sin^2(2\theta_{13})$ (90% C.L.) as a function of the initial source strength and measurement repetitions. Results for both $^{51}$Cr and $^{75}$Se are shown, considering statistical uncertainties only.

**5.2 Sterile neutrinos, mixing angles and very short baseline lengths**

The LENA detector provides unique sensitivity for the possible "massive" sterile neutrino oscillation. According to Eq. (8), the oscillation length $L_{14}$ in case of 3+1 should be rather short, $L_{14} < 1.24$ m for the case of $^{51}$Cr and $m_{14}^2 > 1.5$ eV$^2$. For the 3+2 case of [9], we obtain the $L_{14} = 3.94$ m and $L_{15}=2.13$ m, whereas for values of [10] we have $L_{14} = 2.06$ m and $L_{15}=1.15$ m. Therefore, the oscillation with $\nu_{st}$ could be observed several times within the first 10 m from the source. This opens an excellent possibility for direct oscillometry. It is worthwhile to note here again that oscillation lengths for active and sterile neutrinos, respectively, $L_{23} = 756$ m and $L_{14}$, $L_{15}$ a few meters (all for $^{51}$Cr) are fully disentangled and can be derived independently.

The differential event number d$N$/d$L$ as obtained from Eq. (32) is depicted in Fig. 6, assuming the best-fit reactor antineutrino anomaly (RAA) [7] mixing parameters and a single 55-days run with a $^{51}$Cr source. Statistical uncertainties for a bin width of 1 m are far smaller than the disappearance amplitude.

Like for $\theta_{13}$, we determine the sensitivity to the amplitude $sin^2(2\theta_{14})$ by the integral event number in LENA, using Eq. (36). As the oscillation is fully contained within the detector, the geometrical parameter *g* reaches the maximum value of 50% for *L*<10 m. The resulting sensitivity for RAA parameters is shown in fig. 7 as a function of source strength and repetitions for $^{51}$Cr: The high sensitivity of LENA is clearly demonstrated. Similar sensitivity is expected for $^{75}$Se.

The RAA analysis of [7] only gives a lower limit for $\Delta m^2_{14}$ >1.5 eV$^2$, and therefore an upper limit to the oscillation length $L_{14}$ <1.24 m (for $^{51}$Cr). As long as $L_{14}$ is large compared to the spatial resolution of about 25 cm, a precise determination of this parameter can be expected. Due to the long oscillation baselines accessible in LENA, there is also considerable sensitivity to smaller values of $\Delta m^2_{st}$.

When performing an oscillation analysis based on differential event rates, the length restriction for *L* reaching from 1 to 90 m has to be taken into account. The former is the height of the fiducial volume, the latter corresponds to 4 points with bin widths of the order of the spatial resolution. The corresponding range of mass squared differences is 0.02 eV$^2$ < $\Delta m^2_{st}$ < 2 eV$^2$. Provided sufficiently large mixing, both the mixing angles $\theta_{1i}$ and baseline lengths $L_{1i}$ (i≥4) can be precisely derived from the oscillometry curves. Fig. 8 shows the oscillometry curve for the 3+2 scenario proposed in [9].

The oscillation probability can also be determined by the integral number of ν-e scattering events. The total number does not depend on the spatial resolution. It depends, however, on the fiducial volume, i.e. on the maximum length *L* included in the analysis.

Fig. 9 shows the dependence of the sensitivity for $sin^2 2\theta_{14}$ on the chosen detection volume, based on three different values of $\Delta m^2_{14}$=1.8 eV$^2$, 0.04 eV$^2$ and 0.02 eV$^2$, respectively, a 5 MCi $^{51}$Cr-source, 27.5 days exposure, and a background rate of solar $^7$Be-ν's of 0.5 per day and ton (cf. Sect. 6.2). The optimum value of *L* is not the longest possible; after reaching maximum sensitivity at intermediate values, increasing the volume lowers the sensitivity as due the $1/L^2$ dependence of the signal rate, the signal-to-background ratio deteriorates from this point.

LENA provides a unique possibility to search for the existence of sterile neutrinos. In the integral analysis, a single month-long run with a 5 MCi $^{51}$Cr-source will allow to reach a maximum sensitivity to $sin^2(2\theta_{14})$ on the level of 2x10$^{-3}$, the exact value depending on $L_{14}$. From the differential analysis, $\Delta m^2_{14}$ can be determined over more than two decades of parameter space.

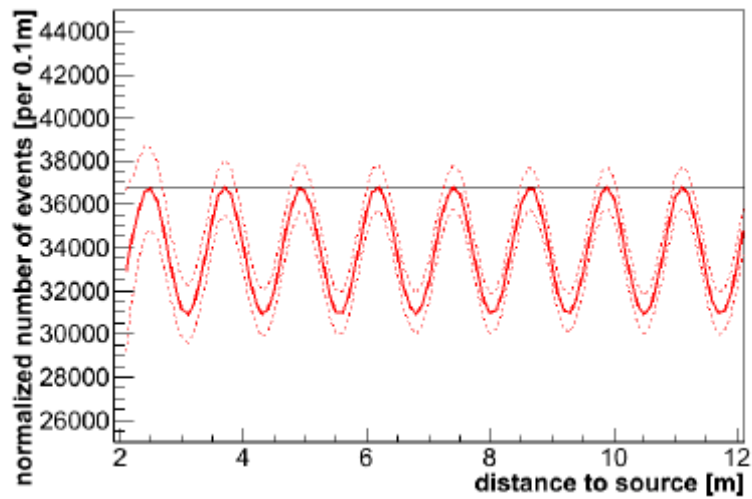

Fig. 6. Differential ν-e scattering event rates for a 55-day run with a 5 MCi $^{51}$Cr-source installed on top of LENA. The first 10 m from the top of the cylinder are shown. The dashed lines indicate the statistical uncertainties (1σ) assuming a bin width of 0.1 m. NA best-fit mixing parameters [7] are used.

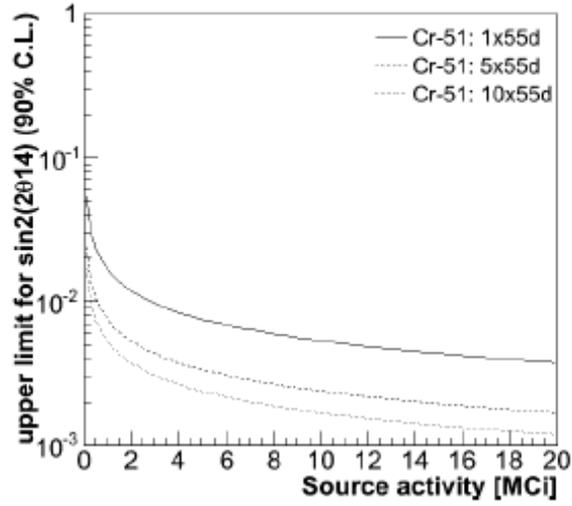

Fig. 7. Upper limits for $sin^2(2\theta_{14})$ (90% C.L.) as a function of the initial source strength and various measurement repetitions. Results for $^{51}$Cr are shown, considering statistical uncertainties only.

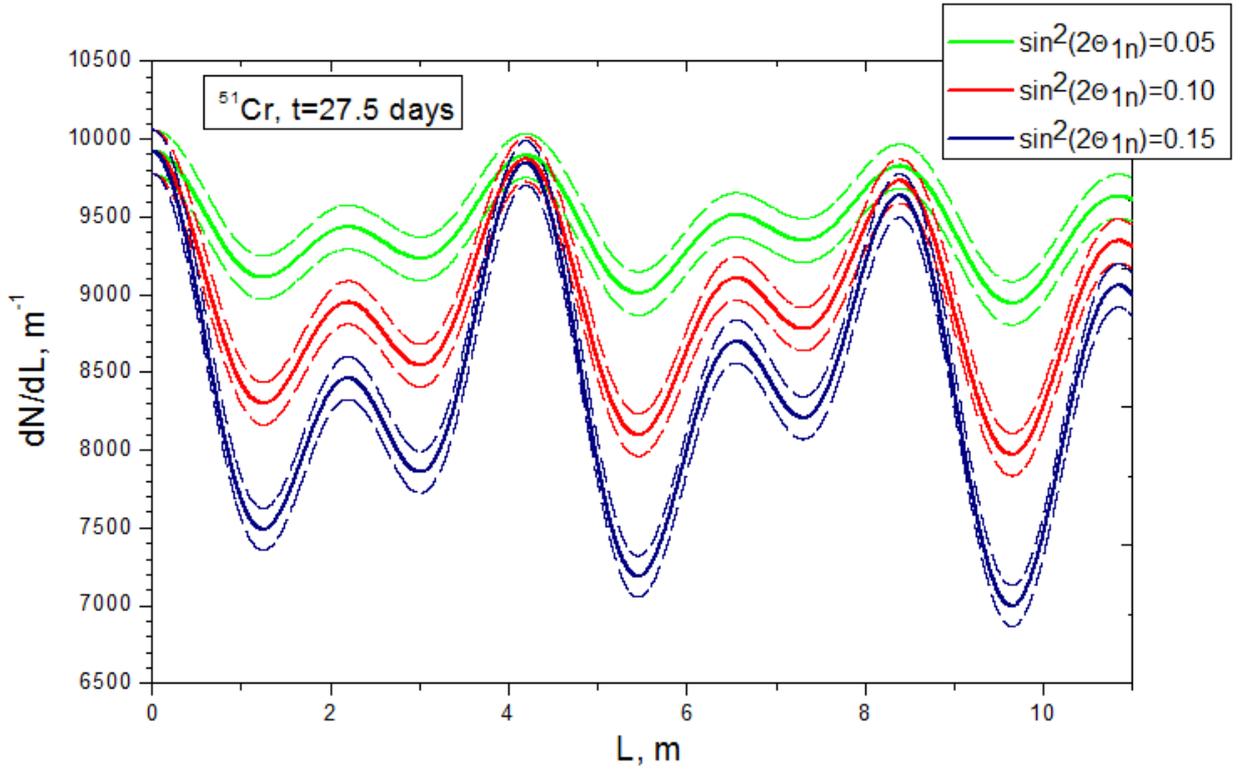

Fig. 8. Oscillometry curves for the case of three active and two sterile neutrinos in the (3+2) scenario with mass parameters proposed in [6]. In the figure top (green), middle (red) and bottom (blue) curves correspond to $sin^2(2\theta_{1n})$ = 0.05; 0.10; 0.15, respectively, with n=4, 5. The dashed lines indicate the statistical uncertainties (1σ). Input parameters are $(T_e)_{th}$ = 200 keV, $R_0$ = 11 m, exposure - 27.5 days, $^{51}$Cr-source intensity – 5 MCi. The background from solar neutrinos is taken from the BOREXINO experiment [26] as 0.5 events/day·t.

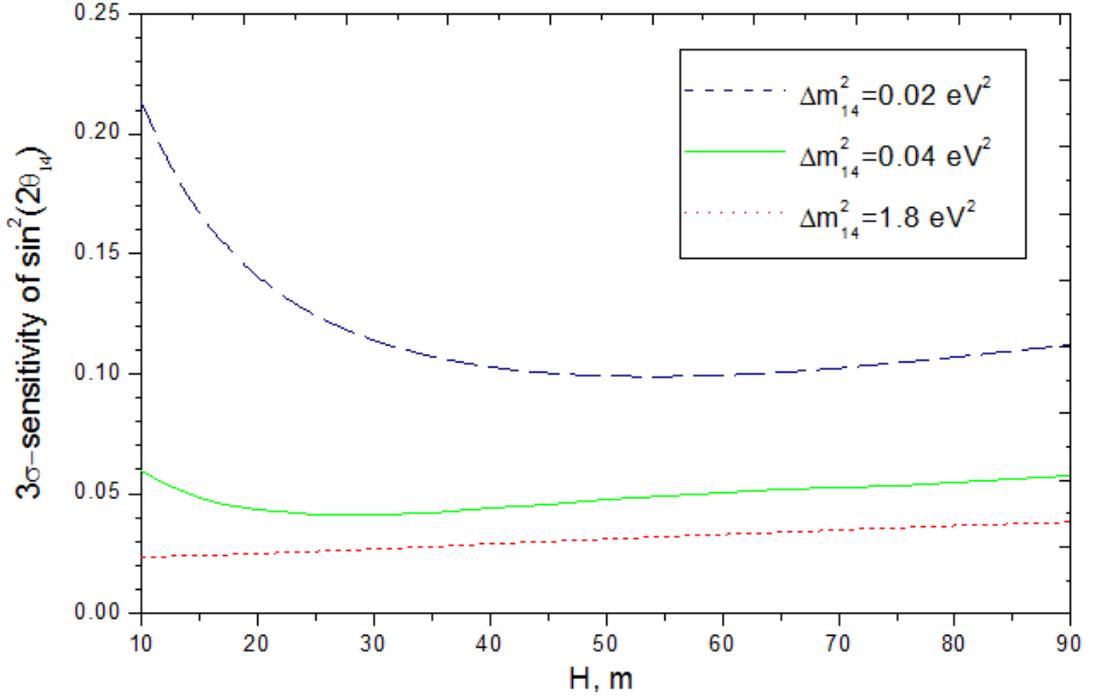

Fig. 9. Sensitivity dependence on the height H of cylinder (with $R_0$=11 m) for the sterile neutrinos with $\Delta m_{14}^2 \approx \Delta m_{24}^2$ = 0.02; 0.04; 1.8 eV$^2$ ($L_{14} \approx$ 93; 46; 1 m, accordingly). Input parameters are $(T_e)_{th}$ = 200 keV, $R_0$ = 11 m, exposure - 27.5 days, $^{51}$Cr-source intensity - 5MCi. The background from solar neutrinos is taken from the BOREXINO experiment [26] as 0.5 events/day·t.

## 6. Expected experimental conditions

An actual experiment will suffer from a number of uncertainties, not discussed previously, which can reduce the sensitivity considered above. Eq. (34) shows the main sources of uncertainties. Among them are the neutrino source initial intensity, the neutrino-electron scattering cross section, the electron density in the liquid scintillator, the absolute value of the neutrino energy, some geometrical uncertainties, the fiducial volume, etc. The influence of these factors is different in the integral and differential analysis of oscillometry curves. In addition, the knowledge of background from solar ($^7$Be) neutrinos and from environment is of paramount importance. The uncertainties from neutrino energy and geometrical parameters have been discussed above. In this section we focus on the two main aspects: the uncertainty of the initial source strength and the subtraction of background events from the signal rate.

### 6.1 Neutrino source intensity

The uncertainty of the initial source intensity will play a role in the analysis of the total number of ν-e events for $\theta_{13}$-determination. However, since $L_{13}$ is known from global analysis, a precise measurement of the neutrino rate in the first 10-20 m of LENA can be exploited as a normalization for the search in the remaining volume, provided the number of events in the near volume exceeds $\sim 10^4$. The uncertainty in source activity will not play a dominant role for the detection of large oscillation amplitudes as in the case of $\theta_{14}$. There is a good experience in precise determination of source intensity accumulated throughout the world. In the calibration campaigns of GALLEX and SAGE experiments that used strong sources based on $^{51}$Cr (≈2 MCi) and 37Ar (≈0.4 MCi), great care was given to an exact determination of the source activity [13], [14]. Various methods were used, ranging from precision measurements of source weight and heat emission to direct counting of decays in aliquots of the sources. The highest accuracy reached for $^{51}$Cr was 0.9% [13], and 0.4% in case of $^{37}$Ar [14].

### 6.2 Background subtraction

Solar neutrinos pose an irreducible background for all oscillometric measurements, constant both over the source exposure and the length of the detector. It is fairly independent of the used artificial neutrino source at the top of cylinder. Solar $^7$Be-neutrinos are be detected at a rate of 0.5 counts per day and ton, featuring a maximum electron recoil energy of 665 keV, only slightly above the recoil spectra maximum of $^{51}$Cr at 560 keV. In addition, radioactive impurities inside the scintillation volume have to be considered: $^{14}$C sets the energy threshold of detection to ≈200 keV, while traces of the nuclides $^{85}$Kr, $^{210}$Po and $^{210}$Bi dissolved in the scintillator will cause background contributions over the whole energy range of the source signals [33]. In the $^7$Be analysis in [26] most of the background contributions are eliminated by pulse-shape discrimination and a spectral fit to the signal region, separating the neutrino recoil shoulder from background spectra. A similar technique could be applied in LENA: the efficiency depending on the achieved photoelectron yield and pulse shaping properties.

This analysis will be aided by the fact that the produced EC-source can be removed from the detector, providing an exact measurement of the background rates. Nevertheless, the $L^2$-decrease of the signal rate means that the background rate will dominate in the far-region of the detector, considerably enhancing the signal rate uncertainties in the analysis of data for $\theta_{13}$-determination. This, however, does not assail the search for sterile neutrinos that mainly concentrates on the parts of the detector closest to the source.

Thus, the feasibility of an oscillometric search will depend on the availability of strong sources, background conditions and the efficiency of spectral separation. These conditions are very crucial mostly for 13-mixing parameters determination.

### 7. Conclusion

Thanks to its relatively low energy detection threshold (200 keV), high overall detection efficiency, considerable length (100 m), and very large detection volume, the proposed liquid-scintillator detector LENA will be exceptionally well suited to perform the determination of neutrino oscillation parameters in both short and very short baselines of 13-, 14- and 15-mixing.

The unique parameters of LENA will compensate for the smallness of the cross-sections for neutrino-electron scattering, which can be used for scanning the number of recoil events along the detector length. As the neutrino flavor changes are within this length, the point-by-point monitoring of events, which we call oscillometry, is feasible in LENA, offering a complementary approach for the determination of neutrino mixing parameters.

We have shown that precise oscillometry is feasible if the neutrinos are monoenergetic and thus the oscillation curve is based on a single oscillation length $L_{1i}$. The optimum neutrino energy is of the order of a few hundred keV, taking into account both the target size and detection threshold of LENA. This energy range fits the half-life of corresponding nuclides, $^{51}$Cr and $^{75}$Se, which have been identified as possible neutrino sources. They can be produced with intensities of a few MCi in high-flux nuclear reactors.

The oscillometry at LENA offers several considerable advantages:
- The oscillation curve is not affected by the matter effects present in long-baseline neutrino beam experiments.
- A considerable part of the short oscillation length $L_{13}$ is within the size of detector. In case of $L_{14}$ and $L_{15}$, the spatial pattern generated by numerous subsequent oscillation minima and maxima is contained inside the detector.
- As the oscillation lengths $L_{13}$ and $L_{14}$ are disentangled, LENA is a versatile tool for a concurrent determination of these lengths as well as mixing angles $\theta_{13}, \theta_{14}$. Both 3+1 and 3+2 scenarios will be accessible.
- A complementary analysis of the differential oscillometry curve as well as the total number of events accumulated in the full fiducial volume allows testing the self-consistency.

- If the monoenergetic neutrino source is transportable, its position outside of the detector can be optimized depending on the tasks desired;
- Due to its large length, LENA can be functionally divided into sub-volumes for different analyses, e.g. a concurrent determination of source strength and oscillation amplitude.
- The background in the detector can be precisely measured if the source is removed.
- Multiple measurements based on various sources of different nuclides can be performed.

Radioactive background and the signal of solar $^7$Be neutrinos can, however, reduce the accurate determination of 13-mixing. This mixing, because of comparably high values of $L_{13}$ for $^{51}$Cr and $^{75}$Se, effectively appears in the far region of detector where the geometrical factor attenuates the effect on the level of strong constant background. Multiple runs with strong sources as well as excellent detector performance and background conditions would be required to reach a sensitivity in $\sin^2 2\theta_{13}$ that is competitive to expected reactor and long-baseline experiments.

LENA offers a large potential to search for the new sterile neutrinos and carefully investigate their properties. This detector can achieve great sensitivity for the sterile neutrinos predicted in [6] and [7]: The best-fit mixing parameters could be conclusively tested by one month run with a 5-MCi $^{51}$Cr-source. Also $L_{1i}$ with i>3 can be determined precisely, providing $\Delta m_{st}^2$ is larger than ~0.01 eV$^2$.

**Acknowledgements**. We would like to thank E. Akhmedov, Y. Azimov, K. Blaum, S. Eliseev, Y. Giomataris, C. Giunti, W. Hampel, T. Kalliokoski, J. Maalampi, A. Marle, T. Schwetz, A. Vassiliev and A. Vorobyov for fruitful discussions.

# References

1. T. Schwetz, H.Tortola and J.W.T.Valle, arXiv:1103.0734v2 [hep-ex] (2011).
2. G.L. Fogli et al., Phys. Rev. **D 84**, 053007 (2011).
3. T2K collaboration, arXiv:1106.2822 [hep-ex] (2011).
4. MINOS collaboration, arXiv:1108.0015 [hep-ex] (2011).
5. SNO-collaboration, arXiv:1109.0763v1 [nucl-ex] (2011).
6. C. Giunti and M. Laveder, Phys. Rev. **D 82**, 053005 (2010), see also arXiv:1012.4356v1 [hep-ph] (2010).
7. G. Mention et al., Phys. Rev. **D 83,** 073006 (2011).
8. C. Guinti and M. Laveder, arXiv:1109.4033v1 [hep-ph] (2011).
9. J. Kopp et al., arXiv:1103.4570v2 [hep-ph] (2011).
10. C. Giunti, arXiv:1106.4479 [hep-ph] (2011).
11. M. Wurm et al., LENA White-Papers, arXiv:1104.5620v2 [astro-ph.IM] (2011).
12. J.D. Vergados and Yu.N. Novikov, Nucl. Phys. **B 839**, 1 (2010).
13. W. Hampel et al., Phys. Lett. **B 420**, 114 (1998).
14. J.N. Abdurashitov et al., J.Phys., Conf. Ser., 39, 284 (2006).
15. A. Ianni, D. Montanino and S. Scioscia, Eur. Phys. J. C8, 609 (1999).
16. J.D. Vergados, Y. Giomataris and Yu.N. Novikov, arXiv: 1103.5307 [hep-ph] (2011).
17. J.D. Vergados, Y. Giomataris and Yu.N. Novikov, Nucl. Phys. **B 854**, 54 (2012).
18. M. Gribier et al., arXiv:1107.2335 [hep-ex] (2011).
19. O. Yasuda, arXiv:1107.4766v2 [hep-ph] (2011).
20. S. Agarwalla et al., arXiv:1105.4984 [hep-ph] (2011).
21. Y. Giomataris and J.D. Vergados, Nucl. Instr. Meth. **A 530**, 330 (2004).
22 L. Oberauer, F. v.Feilitzsch and W. Potzel, Nucl. Phys. **B** (Proc. Suppl) **138**, 108 (2005).
23. P. Vogel and J. Engel, Phys. Rev. **D 39**, 3378 (1989).
24. K. Blaum, Yu.N. Novikov and G. Werth, Contemp. Phys. **51**, 149 (2010).

25. P. Larkins, Atom. Data and Nucl. Data Tables **20**, 311 (1977).
26. C. Asperella et al., Phys. Rev. Lett. **101**, 091302 (2008).
27. M. Gribier et al., Nucl. Instr. and Meth. **A 378**, 233 (1996).
28. http://ie.lbl.gov/ng.html.
29. R. Firestone et al., Table of isotopes, eight ed., vol. 1, John Wiley@sons, inc. N.Y. (1996).
30. V. Barger and D. Marfatia, Phys. Lett. **B 555**, 144 (2003).
31. K. Loo et al., TAUP-conference, Muenchen, Sept. 2011.
32. P. Huber et al., JHEP, 0911:44 (2009).
33. G. Bellini et al., arXiv: 1104.1816 [hep-ex] (2011).